\documentclass[12pt,reqno]{article}

\usepackage{graphicx}	
\usepackage{soul}
\usepackage{amsmath,amssymb,latexsym}	
\usepackage[square,numbers]{natbib}
\usepackage[normalem]{ulem}
\usepackage{notoccite}
\usepackage[totalwidth=460truept,totalheight=600truept]{geometry}

\makeatletter%
\xdef\Equ@width{\the\@tempdima}

\usepackage[T1]{fontenc}
\usepackage{color}

\global\arraycolsep=1truept
\newcommand{\beq}{\begin{equation}}
\newcommand{\eeq}{\end{equation}}
\newcommand{\bea}{\begin{eqnarray}}
\newcommand{\eea}{\end{eqnarray}}

\definecolor{darkred}{rgb}{.8,0,0}

\definecolor{darkblu}{rgb}{0,0,.8}
\newcommand{\cblu}{\color{darkblu}}

\def\d{\displaystyle}

\begin{document}

\vskip 1.4truecm

\begin{center}

\pagestyle{empty}
\vspace{1.2cm}\end{center}
\begin{center}
\LARGE{\bf Hidden conformal symmetry and Entropy of Schwarzschild-deSitter spacetime}\\[12mm] 

\vspace{9mm}
\large
\textsc{D.~Guerrero-Domínguez and  P.~Talavera}

\vspace{8mm}

\footnotesize{
Department of Physics,
Polytechnic University of Catalonia,\\ Diagonal 647,
Barcelona, 08028, E\\
}

\vspace{4mm}
%
{\footnotesize\upshape\ttfamily  daniguerrerodominguez@hotmail.com,  ptalavera@protonmail.ch} \\

\vspace{2mm}
\noindent

\small{\bf Abstract} \\
\end{center}
\begin{center}
\begin{minipage}[h]{\textwidth}
 In this paper we show that the equation of motion of a massless scalar particle near the black hole (BH) horizon in the Schwarzschild-deSitter (SdS) spacetime enjoys a hidden conformal symmetry when the two horizons satisfy a quadratic relation, not related with the Nariai limit. This hidden symmetry is $SL(2,\mathbb{R})\times SO(2)$ instead of $SO(2,1)\times SO(2)$, the expected symmetry of $dS_2\times S^2$. We present a structural analysis of SdS spacetime and compute the leading quantum corrections to the entropy of SdS spacetime in a semi-classical way.
\end{minipage}
\end{center}
\newpage
\setcounter{page}{1}
\pagestyle{plain}

\section{Motivation and Conclusions}
It is known that any spherically symmetric solution of general relativistic vacuum field equations
\bea
G_{\mu\nu}=\Lambda g_{\mu\nu}\,,
\eea
can fit, after a coordinate transformation, the form
\beq
ds^2 = -f(R)\, dT^2 +\frac{1}{f (R)}\, dR^2 + R^2\, d\Omega^2\,,
\label{genNewt}
\eeq
with $d\Omega^2= d\theta^2+ \sin^2(\theta)\,d\varphi^2 $ \cite{Lake:1977ui}.
The  function $f(R)$ in (\ref{genNewt}) for the SdS spacetime is
\beq
f(R)=1-\frac{2 m}{R} -H^2 R^2\,,
\label{fSwchdS}
\eeq 
where $H$ is related to the cosmological constant, $H^2=\Lambda/3$\,.
This class of static, spherically symmetric metrics have the additional 
``Newtonian property'' that the static observers experience a centrifugal constant acceleration just  opposite in value to what in the Newtonian view would 
have been interpreted as the centripetal acceleration of the radial coordinate, for the radial geodesics.

In this paper we examine several aspects of the expanding, $\Lambda>0$, SdS model.  First we review the corresponding location of the horizons and whether these form marginally trapped surfaces. We find that, for an expanding universe, if one deals with a white hole (WH) at the center of the spacetime, the cosmological horizon is trapped and no signal from the boundary can cross it. Next we stress the subtleties of the near extremal limit. Note that in the static coordinates the extremal limit, the so called Nariai black hole \cite{Eune:2012mv} both the event and the cosmological horizon have the same radius because the condition $27m^2H^2=1$ holds. Instead, we will study a case in which the two horizons are close to each other but do not have the same radius. The fact that, near extremality, locally one obtains an $AdS_2\times S_2$ spacetime, suggests that the equation of motion for a massless scalar test particle possesses a hidden symmetry. We find, not surprisingly, that when the event and cosmological horizons are near to each other, but not equal, this is achieved. As a last point we calculate the leading quantum corrections to the entropy in SdS considering that the gravitational background is fixed and the matter corrections provide the largest contribution.
Provided the system is stationary, the mass of the system grows indefinitely with the radial coordinate, see \eqref{misner}, and as a consequence the entropy also shares this behaviour.

\section{Basics on Schwarzschild-deSitter spacetime}

As we shall discuss below (\ref{genNewt}) has coordinate singularities,
but the only essential singularity is at the origin, which is indeed a singularity of the manifold. 
One can check for instance that 
the Kretschmann scalar takes the form 
\begin{equation}
\d K=R_{\mu\nu\alpha\beta}R^{\mu\nu\alpha\beta}=\left(\partial_R^2 f(R)\right)^2 + \left(\frac{2\partial_R f(R)}{R}\right)^2+
\left(\frac{2(1-f(R))}{R^2}\right)^2\,.
\end{equation}

As is customary in order to obtain a regular line element at any of the horizons, we redefine the time coordinate such that the surfaces of constant $t$ are intrinsically flat \cite{Martel:2000rn}
\begin{equation}
   r=R\,,\quad t=T-\epsilon\int\frac{\sqrt{1-f(r)}}{f(r)} dr\,, 
\end{equation}
where the parameter
$\epsilon=1$ describes the 
SdS WH and $\epsilon=-1$ describes the 
SdS BH.
Those coordinates are the so called 
Painlev\'e-Gullstrand coordinates $\left(t,r,\theta, \varphi\right)$.
In term of the latter, \eqref{genNewt} becomes \cite{Shankaranarayanan:2003ya}
\beq
ds^2 = - f(r)\, dt^2 - 2\,\epsilon \,  \sqrt{1 - f(r)}\, dt\,dr + dr^2 + r^2\, d\Omega^2\,.
\label{P-G}
\eeq
With this coordinatization in (\ref{P-G}), the $3$-metric for equal time hypersurfaces is just the 
Euclidean $r$-space,
\beq
dl^2 = dr^2 + r^2\, d\Omega^2\,.
\label{3surfE}
\eeq

\vspace{4mm}

It is instructive to compare the above construction, (\ref{P-G}), to a generic spherically symmetric setting described also in Painlav\'e-Gullstrand coordinates
\begin{equation}
\label{PG}
ds^2=-\sigma(t,r)^2 dt^2+\left(dr-\epsilon\sqrt{\frac{2 M(t,r)}{r}} \sigma(t,r) dt\right)^2+r^2 d\Omega^2\,,
\end{equation}
where $M(t,r)$ is the generalized Misner-Sharp mass function \cite{Misner:1964} and can be interpreted as the quasilocal energy contained in a sphere of radius $r$
at time $t$. In Schwarzschild spacetime this mass function approaches the ADM mass at spatial infinity whereas in radiating spacetimes it reproduces the Bondi mass at future null infinity.  The lapse function, $\sigma(t,r)$ is $r${\cblu -}independent at  spatial asymptotic in the case Schwarzschild spacetime. In our case, matching the functional behavior between
(\ref{PG}) and (\ref{P-G}), one obtains
\begin{eqnarray}
&&\mp \sqrt{\frac{2M(t,r)}{r}}\sigma(t,r)=\mp\sqrt{1-f(r)}\,,\nonumber\\
&&\frac{2M(t,r)}{r}\sigma(t,r)^2-\sigma(t,r)^2=-f(r)\,,
\end{eqnarray}
which solution  is
\begin{equation}
\label{misner}
\sigma(t,r)=1\,,\quad M(r)= m+\frac{H^2}{2} r^3\,.
\end{equation}
In terms of this local, Tolman-like, mass function \cite{Gibbons:1976ue,Bhattacharya:2013tq}, the metric \eqref{genNewt} can be written as a Schwarzschild-like spacetime
\bea
ds^2 = -\left(1-\frac{2M(R)}{R}\right) \, dT^2 +\frac{1}{\left(1-\frac{2M(R)}{R}\right)}\, dR^2 + R^2\, d\Omega^2\,.
\label{genlocal}
\eea
At this point 
 the divergence of the generalized Misner-Sharp function at spatial infinity has the following physical interpretation: as
the energy-momentum tensor  contains a constant pressure term, its integration over the full space grows as the size of the
radius of the sphere increases. The previous definition, (\ref{misner}), only requires the spacetime to be spherically symmetric and neither be static nor asymptotically flat.

\section{Structural analysis} \label{sec:Structural analysis}

Time{\cblu -}dependent solutions of general relativity may behave radically different from their stationary counterparts. Some common physical properties which rely on time independence 
and asymptotic flatness do not hold for more general, dynamical, settings. By Birkhoff's theorem, any spherically symmetric solution to the vacuum Einstein equations can be cast in the form (\ref{genNewt}), with $m$ being constant.  

\subsection{General structure: Apparent Horizons}

Apparent horizons (AH) are defined as surfaces where congruences of geodesic null vectors change the focusing properties. If we expand the spherical congruences
along radial null directions, the AH  satisfy
\begin{equation}
\Theta_{-}\ \Theta_{+}=0\,,\quad \Theta_\pm=\epsilon\sqrt{1-f(r)}\pm 1\,,
\end{equation}
that boils down to 
\begin{equation}
\label{ho}
f(r)=0\,,\quad H^2 r^3= r - 2 m\,,
\end{equation}
for \eqref{fSwchdS}.  The above cubic equation has only real solutions for $3 H\le \frac{1}{\sqrt{3} m} $. 
If we think of the horizons as the intersection of the two curves in \eqref{ho}, it becomes clear that there can be at most two admissible real positive roots \cite{Nickalls:1993}, see Fig. 1,
\begin{eqnarray}
\label{horizons}
r_{+}= \frac{2}{\sqrt{3} H} \cos \theta\,\,, \quad r_{-}= \frac{2}{\sqrt{3} H} \cos \left(\theta+\frac{4\pi}{3}\right)\,,
\end{eqnarray}
 and one negative
\begin{eqnarray}
r_{--}= \frac{2}{\sqrt{3} H} \cos \left(\theta+\frac{2\pi}{3}\right)\,,
\end{eqnarray}
with $\cos 3\theta =- 3\sqrt{3} H m $ and the ordering $0<2m<r_{-}<3m<r_{+}$ \cite{Lake:1977ui, PhysRevD.15.2738}. 
Furthermore, since \eqref{fSwchdS} has no linear term in $r$, the three roots fulfill the relation $r_{--}+r_-+r_+=0$. 
Out of these positive roots, one can define the three regions, or basic buildings blocks, for the construction of the Penrose-Carter diagram \cite{walker:1970}.

In the region between the two horizons $\Theta_{-}< 0 < \Theta_{+}$ and it constitutes the regular region. Outside this,
$\Theta_{-}  \Theta_{+}>0$,
we shall deal with future trapped regions, $\Theta_\pm <0$, for BHs and with past trapped regions, 
$\Theta_\pm >0$, for WHs.

\vspace{0.5cm}

\begin{figure}
\begin{center}
\includegraphics[width=0.81\linewidth]{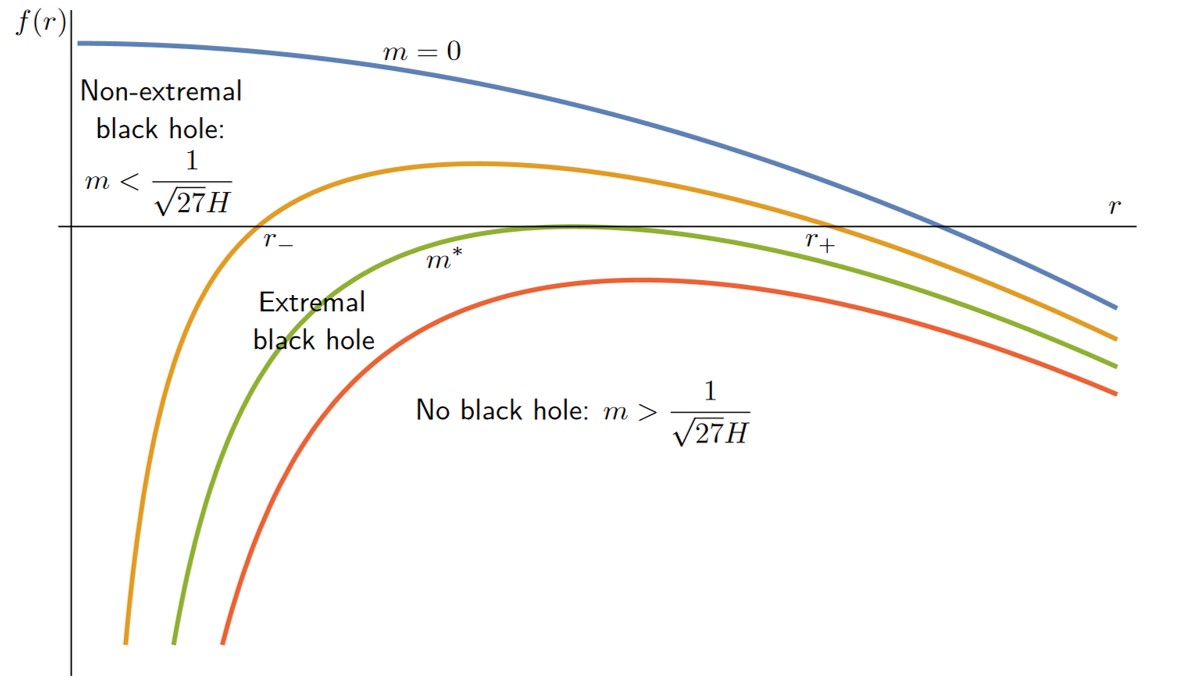}\caption{Representation of the factor $f(r)$ for different values of $m$. The blue line corresponds to the de Sitter space, the yellow line to the SdS space with two horizons, the green line to the extremal case in which both horizons coincide. The red line corresponds to the case where $m$ is so large that neither we have event horizon nor cosmological horizon.}
\end{center}
\end{figure}

\subsection{Trapped Surfaces}

It is widely known that surface gravity is a concept well founded only for static or stationary spacetimes \cite{Wald:1984rg}. This is so because generically, under such circumstances, there exists
a global time translational Killing vector field that becomes null on the event horizon. For dynamical situations the r\^ole of the Killing vector does no longer exist and
local definitions of horizons do not necessarily coincide with the location of the event horizon. 
This can be a relevant issue since it is the role of the AH which seems to matter during Hawking emission. Here we explore some known solutions.

For a spacelike surface there are two independent normal vector fields that can be chosen future-pointing and null everywhere with 
\begin{equation}
\label{condi}
k_\mu^\pm k^{\pm\mu}=0\,,\quad k_\mu^\pm k^{\mp\mu}=-1\,.
\end{equation}

The system of coordinates (\ref{P-G}) is regular at future horizons. The outgoing and ingoing radial null vectors, orthogonal to the constant $(t,r)$ surfaces are given respectively by
\begin{eqnarray}
\label{nulls}
k^{+\mu}=\frac{\partial}{\partial t}+ \left( 1-\epsilon \sqrt{1-f(r)}\right)\frac{\partial}{\partial r} \,,\quad
k^{-\mu}=\frac{1}{2} \frac{\partial}{\partial t}- \frac{1}{2} \left(1+\epsilon\sqrt{1-f(r)} \right)\frac{\partial}{\partial r}  \,, 
\end{eqnarray}
with the normalization  $k^+_\nu\ k^{-\nu}= -1\,.$ Notice that these vectors grow unbounded at the spatial infinity. In addition for the AdS case
they satisfy the null energy condition
\begin{equation}
G^\mu_\nu k^-_\mu k^{-\nu} = G^\mu_\nu k^+_\mu k^{+\nu} = 0\,,
\end{equation}
being $G^\mu_\nu$ the Einstein tensor. The above expressions asserts that there is \sout{not} {\cblu no} energy flux across the instantaneous location of the horizon.
The expansions of the null vectors  are
\begin{eqnarray}
\label{thetas}
\theta_+(r)&=&q^{\mu\nu}\nabla_\mu k^+_\nu= \frac{2}{r} \left(1+\epsilon \sqrt{1-f(r)} \right)\,,\nonumber \\
\theta_-(r)&=&q^{\mu\nu}\nabla_\mu k^-_\nu=-\frac{1}{r} \left(1-\epsilon \sqrt{1-f(r)} \right)\,,
\end{eqnarray}
where $q^{\mu\nu}$ is the projector onto the  two-surface normal to $k^{+\mu}$ and $k^{-\mu}$.

Using them we can calculate the scalar $\kappa$, i. e. the point where a purported  global time translational Killing vector field becomes null.
There are many definitions for it. 
One can pick out any of such constructions keeping in mind two well known issues: 
\begin{enumerate}
\item[ {\sl i)}] In the static case all reduce to the static Killing definition by construction, see for instance \cite{Fodor:1996rf}.
\item [ {\sl ii)}] The location of $\kappa$ is observer{\cblu -}independent. 
 \end{enumerate}
In our case to construct $\kappa$ we rely on the trace of the shape tensor, $\vec{H}$, for the codimension two surface perpendicular to the sphere. We follow
  \cite{Senovilla:2011fk} slightly adapted to allow the simultaneous treatment of BH and WH. We are after the different regions of the spacetime where  $\vec{H}$
  changes its causal character
 \begin{equation}
\kappa=\epsilon g^{ab}H_a\ H_b\vert_{\rm on\ the \ horizon}\, \quad{\rm with}\quad  \{a,b\}=\{r,t\}\,\quad {\rm and}\quad \vec{H}=-\theta_-(r)\vec{k}^+-\theta_+(r)\vec{k}^-\,.
\end{equation}
Inserting (\ref{nulls}) and (\ref{thetas}) in the previous expression \cite{Bousso:1997wi}
one obtains
\begin{equation}
\label{kappa}
\kappa=\epsilon \frac{4}{r^2} f(r)\,,
\end{equation}
as must be the case for a spacetime with spherical symmetry \cite{Senovilla:2011fk}.
The vanishing of (\ref{kappa}) reduces to (\ref{ho}) 
 and either for BHs or WH both horizons, (\ref{horizons}), are marginally trapped.
 For BH,  $\kappa<0$ in between $r_-<r<r_+$ and the surface is untrapped. While for WH, $\kappa>0$ and the surface is trapped.

\subsection{Near-extremal limit}

We now pay a closer look to the particular, extremal case $m=\frac{1}{3 \sqrt{3} H}$, 
the so called Nariai metric \cite{1950SRToh..34..160N,Nariai1951OnAN}, in which the inner and outer horizons merge at $r_+=r_-$ in the static patch.
Defining the extremal limit parameter $\delta=\frac{r_+-r_-}{r_0}\ll 1$ and the near horizon limit $\epsilon=\frac{r-r_-}{r_0}\ll 1$ with $r_0$ being the near horizon radius in the extremal case, (\ref{genNewt}) can be approximated, near the horizon, in terms of these parameters as 
\beq\label{nearextremalmetric}
ds^2 = -H^2 r_0^2 \epsilon \frac{(\epsilon+\Lambda)(\epsilon+\delta)}{(1+\epsilon)}\, dT^2 +\frac{(1+\epsilon)}{H^2 \epsilon(\epsilon+\Lambda)(\epsilon+\delta)}\, d\epsilon^2 + r_0^2\, d\Omega^2\,,\quad
 \Lambda=: \frac{r_0+r_{--}}{r_0}\,,
\eeq
that leads to different limiting geometries depending on the order in which the limits
 are taken\footnote{The extremal limit corresponds to $\delta\to 0$ whether the near horizon limit is $\epsilon \to 0$.}.
If in \eqref{nearextremalmetric} one takes first the ordering: near horizon limit and afterwards the extremal limit (\ref{genNewt}) reduces to
\begin{equation}
ds^2 = - H^2 r_0^2 \epsilon\delta \Lambda dt^2 + \frac{1}{ H^2 \epsilon\delta \Lambda} d\epsilon^2 + r_0^2 d\Omega^2\,,
\end{equation}
that is singular in the near horizon limit. If contrariwise, we commute the limits,
first performing the extremal limit and subsequently the near horizon limit
one obtains locally dS$_2\times S_2$ \cite{Ginsparg:1982rs}
\begin{equation}
ds^2 = - H^2 r_0^2 \epsilon^2 \Lambda dt^2 + \frac{1}{ H^2 \epsilon^2 \Lambda} d\epsilon^2 + r_0^2 d\Omega^2\,.
\label{dss2}
\end{equation}
Thus, the near horizon limit of the extremal geometry is not the same as the extremal case of the near horizon of the generic non-extremal initial spacetime. 
This near extremal-limit behaviour, (\ref{dss2}), is also found in the Kerr metric \cite{Pradhan:2012yx}. Althought SdS is obviously neither flat nor AdS as the cases studied in \cite{Kunduri:2007vf}, \eqref{dss2} suggests that the SdS space have a hidden symmetry $SL(2,\mathbb{R})\times U(1)$ at the extremal configuration \cite{1950SRToh..34..160N,Nariai1951OnAN,Ginsparg:1982rs}. This was already the case for a rotating Nariai geometry \cite{Anninos:2009yc}.

\section{Enhanced scalar symmetry near extremality}
In the following we show that the massless scalar wave function at low frequencies also enjoys
 a hidden $SL(2,\mathbb{R})\times SO(2)$ symmetry instead of the $SO(2,1)\times SO(2)$ symmetry of (\ref{dss2}), when the horizons $r_+$ and $r_-$ fulfill some quadratic relationship. Our starting point is 
\begin{equation}
\frac{1}{\sqrt -g} \partial_\mu \left( \sqrt{-g} g^{\mu\nu}\partial_\nu\right) \Phi=0\,.
\label{quadratelo}
\end{equation}
To start with, we decompose the wave function
\beq
\Phi(t , r,\theta,\phi)=e^{-i\omega t} R(r) Y^l_m(\theta,\phi)\,,
\eeq
where $Y^l_m(\theta,\phi)$ stand for the $S^2$ spherical harmonics, $\nabla_{S^2}Y^l_m(\theta,\phi)=l(l+1)Y^l_m(\theta,\phi)$ and $R(r)$ stands for the radial part of the wave function.

We stress that we do not want to find the explicit solution of \eqref{quadratelo} but just to figure out its symmetries near one of the horizons, let's say $r_-$.
As is customary, we rewrite the factor $f(r)=G(r) \frac{\Delta(r)}{r^2}$, where
the function $G(r)=-H^2(r-r_{--})(r-r_+)$  does not vanish in the near inner horizon limit while
$\Delta(r)=r(r- r_-)$ vanishes in the same limit.
Near the inner horizon region and at low frequency limit,  equation \eqref{quadratelo} becomes
\beq
G(r)\Delta(r) \partial_r^2 R(r)+\partial_r\left[G(r)\Delta(r)\right]\partial_r R(r)+\left(\frac{r_-^4 \omega^2}{G(r)\Delta(r)}-l(l+1)\right)R(r)=0\,.
\label{eqdif1}
\eeq

As is already known, \cite{Bertini:2011ga}, near the Schwarzschild BH horizon, equation \eqref{quadratelo} has an $SL(2,\mathbb{R})$ symmetry. It would be surprising if this still holds when we enlarge the model with a second parameter, the cosmological constant. We shall see in the following when this is indeed the case.

We define the set of global vectors fields
\bea
\label{sl2}
 H_1&=&i e^{t/(2r_-)}(\Delta G)^{1/2} \left\{
 \left(\frac{ \left(\partial_r(\Delta G)\right)^2-4\Delta G}{\left(\partial_r(\Delta G)\right)^2-2\Delta G \partial^2_r\left(\Delta G\right)} \right) \partial_r - r_- \partial_r\left(\ln( G\Delta)  \right)\partial_t \frac{}{}\right\}\,, \nonumber \\
 H_0&=&-2 i r_-\partial_t\,,\nonumber\\
H_{-1}&=&-i e^{-t/(2r_-)}(\Delta G)^{1/2} \left\{ \left(\frac{ \left(\partial_r(\Delta G)\right)^2-4\Delta G}{\left(\partial_r(\Delta G)\right)^2-2\Delta G \partial^2_r\left(\Delta G\right)} \right) \partial_r + r_- \partial_r\left(\ln( G\Delta)\right)  \partial_t \frac{}{}\right\}\,.\nonumber\\
\eea
These three generators obey the global $SL(2,\mathbb{R})$ commutations relations 
\beq
[H_0,H_{\pm 1}]=\mp i H_{\pm 1}\,,\quad [H_1,H_{-1}]=2iH_0\,.
\eeq
 For $G(r)=1$ \eqref{sl2} reduces to the Schwarzschild case \cite{Bertini:2011ga}, while the de Sitter solution is
recovered setting $G(r)=1$ and $m=0$ \cite{Anninos:2011af}.

The corresponding Casimir, in the near horizon limit, reads as
\begin{eqnarray}
\mathcal{H}^2&=& -H_0^2+\frac{1}{2}\left(H_1H_{-1}+H_{-1}H_1\right)
\nonumber \\
&=&
G(r)\Delta(r) \partial_r^2 R(r)+\partial_r\left[G(r)\Delta(r)\right]\partial_r R(r)+
\frac{r_-^4 \omega^2G(r)}{\Delta(r)}R(r)\,.
\end{eqnarray}
This would correspond to \eqref{eqdif1} in the near horizon limit, and so the Klein-Gordon equation \eqref{quadratelo} would read as $\mathcal{H}^2\Phi=l(l+1)\Phi$, if and only if
\beq\label{condition}
G(r_-)^2=1\,,
\eeq
that, after imposing $r_+>r_-$, reduces to the simple quadratic relation between the location of the two positive horizons $G(r_-)=1$. Then, the condition $G(r_-)=1$ has the relation
\bea
\label{constraint}
r_+=\frac{1}{2}\left(-r_-+\sqrt{9r_-^2+\frac{4}{H^2}}\right)
\eea
as solution.
 Notice that, generically, the enhancement of symmetry for the Klein-Gordon equation \eqref{quadratelo} to $SL(2,\mathbb{R})$ is not achieved at the extremal limit of \eqref{genNewt}. This finding runs along the suggestions in \cite{Stotyn:2015qva}. There are in order two considerations. First, the difference $r_+-r_-$ fulfilling \eqref{constraint} vanishes when one considers large inner horizons, or what is equivalent small masses $m$ in \eqref{genNewt}. 
Then only if $r_-\gg$ the isometries of \eqref{genNewt} in the Nariai limit match with the symmetries of \eqref{quadratelo}. 
Second, if one approaches the Schwarzschild case, $H\to 0$, the outer horizon is pushed to the boundary.

\section{Leading quantum correction to the entropy}
\label{leading}

The main goal of this section is to calculate the quantum corrections to the SdS space and BH entropy using an Euclidean approach \cite{Solodukhin:1994yz}. 
By and large, we will be rederiving old and widely known techniques, but the exercise will be an instructive illustration of the workings of the manifold with a conical singularity in our context. 
New studies in models like the presented in \eqref{genNewt},
\eqref{fSwchdS}
in four dimensions \cite{Banihashemi:2022jys}, and its reduced 2-dimensions versions \cite{Svesko:2022txo}, have been constructed using the notion of causal diamond
obtaining a further understanding of the Smarr relations and the thermodynamics first law.

We briefly review the argument and apply it to our particular set up.
The starting point is the canonical ensemble of a composed system of gravitational and matter fields at finite temperature in the stationary phase. If one neglects the graviton contributions and only quantizes the matter sector, the one-loop approximation for the partition function is given by
\beq
\label{genfunc}
Z(\beta)={\rm exp}\left[-I_{\rm gr}(g,\beta)-\frac{1}{2}\ln\det\Delta_g\right]\,,
\eeq
where $\ln\det\Delta_g$ is the formal integration over the matter content at a fixed gravitational background.
The periodic identification of the time coordinate $\phi$ with period $2\pi \beta$, i.e. $\phi=i T/\beta$, gives a topology $C_2\times S_2 $ for (\ref{genNewt}).
The general expression for the thermodynamic entropy is
\beq
\label{entropy}
S=\left(-\beta\partial_\beta+1\right)\ln Z(\beta)\,.
\eeq
Inserting (\ref{genfunc}) and evaluating at any of the horizons, the
entropy splits into two contributions
\bea
\label{entropiess}
S_{\rm BH}=\left. S\right|_{\beta\to \beta_{\rm h}}=S_{\rm BH}^{\rm class}+S_{\rm BH}^{q}\,,
\eea
the former takes into account the classical gravitational term and will
leads to the Bekenstein-Hawking term
while the second stands for the matter leading quantum corrections
\beq
\label{entropies}
S_{\rm BH}^{\rm class}= \left(\beta\partial_\beta-1\right)
I_{\rm gr}(g,\beta)_{\beta\to \beta_{\rm h}}\,, \quad
S_{\rm BH}^{q}=\left(\beta\partial_\beta-1\right) {\cal G}_{\rm eff}(\Delta_g)\,.
\eeq

Although we deal with a $3+1$ manifold the fact that it is not fibrated simplifies substantially the description. In fact, the BH emission is essentially a 2-d effect and the $S_2$ acts as a spectator without any significant role.
If one rewrites (\ref{genNewt}) in terms of the compactified time, $\phi$, and the tortoise coordinate $\rho=\int \frac{dr}{\sqrt{f(r)}}$,
the $C_2\times S_2$ underlying structure becomes manifest
\begin{equation}
    ds^2=\beta^2f(\rho)d\phi^2+d\rho^2+r^2(\rho)d\Omega^2\,.
\label{genCon}
\end{equation}
Notice that for general $\beta$ there is a deficit angle $\delta=2\pi(1-\alpha)$, where $\alpha$ is $\alpha=\beta/\beta_{\rm h}$.

One can choose any point in the $r$ coordinate and define the coordinate $\rho$ such that $\rho=0$ at that point. It is then usually useful to expand the functions $f(\rho)$ and $r(\rho)$ in power series of $\rho$ around that point.
In particular, near the inner (outer) horizon, $r=r_{\rm h}(=r_\pm)$, we obtain
\begin{equation}
    f(\rho)\approx \frac{1}{\beta_{\rm h}^2}\rho^2+\frac{\partial_r^2f(r_{\rm h})}{6\beta_{\rm h}^2}\rho^4\,,
\quad
r(\rho)\approx r_{\rm h}+\frac{\partial_r f(r_{\rm h})}{4}\rho^2\,,
\end{equation}
where $\beta$ is related to the temperature at the horizon by $\beta_{\rm h} T_{\rm h}=1/2\pi$ and thus
\begin{equation}
\label{temp}
    T_\pm=\left|\frac{\partial_r f(r)}{4\pi}\right|_{r=r_\pm}=\frac{H^2}{4\pi r_\pm}(r_+-r_-)(2r_\pm+r_\mp)\,,\quad {\rm with}\quad T_- > T_+\,.
\end{equation}

A few words regarding the $C_2$ piece are in order to proceed further \cite{Fursaev:1995ef}:
In 2d the metric of the cone, 
\beq
\label{2dcone}
ds^2=\alpha^2\rho^2 d\phi^2+d\rho^2\,,
\eeq 
is singular just at the tip. To smooth out this singularity we embed the structure in 3d Euclidean space by the transformation $x=\alpha \rho\cos\phi\,,y=\alpha \rho\sin\phi\,,z=\sqrt{1-\alpha^2}\sqrt{\rho^2+a^2}$. This leads to
\begin{equation}
\label{regcone}
ds^2=\alpha^2\rho^2d\phi^2+\frac{\rho^2+a^2\alpha^2}{\rho^2+a^2}d\rho^2\,.
\end{equation}
In the limit $a\to 0$, (\ref{regcone}) coincides with (\ref{2dcone}). In what follows we enlarge the spacetime (\ref{regcone}) to $C_2\times \mathbb{R}^2$
\begin{equation}
\label{regcone2}
ds^2=\alpha^2\rho^2d\phi^2+\frac{\rho^2+a^2\alpha^2}{\rho^2+a^2}d\rho^2+dx^2+dy^2\,.
\end{equation}

One can now deal with (\ref{regcone2}) and calculate the scalar curvature (in the distribution sense) and afterwards send the regulator to zero, obtaining
\begin{equation}
R_{\rm con}=\frac{2(1-\alpha)}{\alpha}\delta(\rho)\,,
\label{Rcon}
\end{equation}
where $\delta(\rho)$ is defined\footnote{More explicitly: $\delta_s(\rho)=\frac{s^2(\alpha+1)}{\alpha(1+s^2\rho^2)^2}$ with $s=(a\alpha)^{-1}$ and $\delta(\rho)= \lim_{s\to\infty}\delta_s(\rho)\,.$} with respect to the measure
\beq
\int_0^\infty\delta(\rho)\rho d\rho=1\,.
\eeq
With a little abuse of the language, it is easy to verify the following behaviour  
\bea
R_{\rm con}\sim R_{\rm con\, \mu\nu}\sim R_{\rm con\,\mu\nu \alpha\beta}\sim\delta(r)\,,
\eea
that will provide useful in the following.

Bearing in mind the cone regularization \eqref{regcone}, the metric \eqref{genCon} is finally recast as
\begin{equation}\label{regularized metric}
ds^2=\beta^2f(\rho)d\phi^2+\frac{\rho^2+a^2\alpha^2}{\rho^2+a^2}d\rho^2+r^2(\rho)d\Omega^2\,.
\end{equation}
With some work one concludes that the Riemann tensor of the product space is simply the sum of the conical and the regular part (which is non-singular in the limit $a\to 0$)
\begin{equation}
    R^\mu_{\nu\alpha\beta}=R^{\mu}_{\text{reg } \nu\alpha\beta}+R^{\mu}_{\text{con } \nu\alpha\beta}\,. \label{Riemann}
\end{equation}
The only non-vanishing piece arising from the cone is
\bea
R^{\phi}_{\text{con } \rho\phi\rho}=\frac{a^2(1-\alpha^2)}{(\rho^2+a^2)(\rho^2+a^2\alpha^2)}\,.
\eea
Concerning the regular part, the Riemann tensor behave as 
\bea
R^{\phi}_{\text{reg } \rho\phi\rho}(\rho=0)=-\left.\frac{\partial_r^2f(r)}{2}\right|_{r=r_{\rm h}}\,. 
\eea
Note that for the Schwarzschild case we recover $R^{\phi}_{\text{reg } \rho\phi\rho}(\rho=0)=4/\beta_{\rm h}^2$ computed in \cite{Solodukhin:1994yz}.

\subsection{Effective action}

The next task is to obtain the expression for ${\cal G}_{\rm eff}(\Delta_g)$ in (\ref{entropies}). 
This consists of a divergent and a finite piece. The latter contains essentially all counter-terms which preserve parametrization invariance and cancel the divergences of the former and/or enables to write any observable in a regulator{\cblu -}independent way. Thus adding the scalar field as a source
\bea
I_{\rm matter}= \frac{1}{2}\int d^4x \sqrt{-g}\,\nabla_\mu\varphi\nabla^\mu\varphi\,,
\eea
the divergent part of the one-loop effective action in four dimensions is \cite{DeWitt:1975ys}
\begin{equation}\label{effAct}
    \mathcal{G}_{\rm inf}=-\frac{1}{32\pi^2}\left(B_0L^4+B_2L^2+B_4\ln{\left(\frac{L}{\mu}\right)^2}\right)\,,
\end{equation}
where $L$ is the ultraviolet cut-off, $\mu$ is a dimensional factor and the $B_k$ coefficients take the form
\begin{equation}\label{B Soloduhkin}
    \begin{split}
        B_0&=\frac{1}{2}\int d^4x\sqrt{-g}\,,\\
    B_2&=\frac{1}{6}\int d^4x \sqrt{-g} R\,, \\
 B_4&=\int d^4x \sqrt{-g} \left( \frac{1}{74}R^2-\frac{1}{180}R_{\mu\nu}R^{\mu\nu}+\frac{1}{180}R^{\mu\nu}_{\ \ \alpha\beta}R_{\mu\nu}^{\ \ \alpha\beta}+\frac{1}{30}\Box R\right)\,.
    \end{split}
\end{equation}
Explicitly using \eqref{Riemann} and subsequent expressions, one gets
\begin{equation}\label{B0B2}
    B_0=\frac{1}{2}\alpha_{\rm h}V\,,
\quad
    B_2=\frac{2}{3}\pi (1-\alpha_{\rm h})A_{\rm h}\,,
\end{equation}
where $V$ is the volume of the spacetime \eqref{genCon} if $\alpha_{\rm h}=1$.
Finally, in a similar way, we can compute $B_4$ in the limit $a\to0$\footnote{ Note that the term  with $T(\alpha_{\rm h})$ contains an $a^2$ in the denominator. However, in the limit $a\to 0$ we do not need to worry about this divergence because, since it is multiplied by $(\alpha_{\rm h}-1)^2$, it will not contribute to the quantum corrections to the entropy when $\alpha_{\rm h}=1$. }
\begin{equation}\label{B4}
        B_4
        =2\pi A_{\rm h} (\alpha_{\rm h}-1)\left\{ \frac{1}{15\alpha_{\rm h}^2} \left.\partial_r^2f(r)\right|_{r_{\rm h}}+\frac{1}{5 r_{\rm h} \alpha_{\rm h}^2} \left.\partial_rf(r)\right|_{r_{\rm h}}- \frac{1} {9 r_{\rm h}^2}
        +\frac{(\alpha_{\rm h}-1)}{15} \frac{T(\alpha_{\rm h})}{a^2}\right\}\,,
\end{equation}
where \begin{equation}
    T(\alpha_{\rm h})=\frac{(1+\alpha_{\rm h})}{\alpha_{\rm h}^4}
\end{equation}
is a smooth function at $\alpha_{\rm h}=1$ such that $T(1)=2$.
 
With all these ingredients, the effective action \eqref{effAct} is finally
\begin{equation}\label{final effact}
\begin{split}
   \mathcal{G}_{\rm inf}=&-\frac{1}{64\pi^2}\alpha_{\rm h} V L^4+\frac{1}{48\pi} (\alpha_{\rm h}-1)A_{\rm h}L^2 
     \\
     &
     -\frac{1}{16\pi}(\alpha_{\rm h}-1)A_{\rm h} \left\{
     \frac{\left.\partial_r^2f(r)\right|_{r_{\rm h}}}{15\alpha_{\rm h}^2}
     +\frac{\left.\partial_rf(r)\right|_{r_{\rm h}}}{5 r_{\rm h}\alpha_{\rm h}^2} 
    -\frac{1}{9 r_{\rm h}^2} +\frac{(\alpha_{\rm h}-1)}{15} \frac{T(\alpha_{\rm h})}{a^2}\right\} \ln{\left(\frac{L}{\mu}\right)}\,.
  \end{split}
\end{equation}

\subsection{Entropy}
Before quoting our result we pause to frame the findings. The BH or WH entropy in (\ref{entropiess}) is sourced by two terms \cite{Bousso:1997wi}. 
There is a first
contribution due to the BH evaporation. This is proportional to the d.o.f. located at the inner horizon \cite{Bekenstein:1973ur,Hawking:1971tu}. A second contribution is coming from the radiation of the cosmological horizon and is due to the presence of $\Lambda$ \cite{PhysRevD.15.2738}. This latter is proportional to the d.o.f. of the outer horizon. Using Wald's entropy \cite{Wald:1993nt} one finds
\cite{Eune:2012mv}
\bea
\label{centropy}
S^{\rm class}= \frac{A}{4\mathsf{G}_4}=S_{r_-}^{\rm class}+S_{r_+}^{\rm class}\,,
\eea
where $A=A_{r_-}+A_{r_+}$ is the total area  and $\mathsf{G}_4$ is Newton's constant in 4d. 

With that in mind we can now compute, using \eqref{entropy} and \eqref{final effact}, the matter quantum corrections to the entropy
contained in an generic sphere of radius $r_-<r<r_+$
\begin{equation}
\label{entropycorrection}
\begin{split}
     \delta \bar{S}^{\rm q}
     =\frac{1}{48\pi} \frac{A_{r}}{\epsilon^2}+\frac{A_{r}}{4}\left\{    \frac{1}{36\pi}\frac{1}{r^2}-\frac{1}{60\pi}\partial_r^2f(r)-\frac{1}{20\pi r}\partial_rf(r)\right\}\ln{\left(\frac{\Sigma}{\epsilon}\right)}\,,
\end{split}
\end{equation}
where $\epsilon:=L^{-1}$ is the ultraviolet cut off and $\Sigma$ is the size of a box in order to eliminate infrared problems.

As is well-known, the matter corrections to the entropy have a
short distance area-like divergence coming from $B_2$, first term in (\ref{entropycorrection}), \cite{1993PhRvL..71..666S}. When this term is added together with (\ref{centropy}) the net effect is to renormalize the Newton constant \cite{Susskind:1994sm}
\begin{equation}
\label{entropyfinal}
\begin{split}
     S
       =\frac{A}{4\mathsf{G}_4^{\rm renor}}+\delta S^{\rm q}_{\rm }\,,
\end{split}
\end{equation}
 with $\delta S^{\rm q}$ being, for the SdS metric,
\begin{equation}\label{qentropy}
\delta S^{\rm q}=\frac{A_{r}}{4}\left\{    \frac{1}{36\pi}\frac{1}{r^2}-\frac{m}{30\pi r^3}+\frac{2 H^2}{15\pi }\right\}\ln{\left(\frac{\Sigma}{\epsilon}\right)}\,.
\end{equation}
If we dubbed as “classical” entropy the first term in \eqref{entropyfinal}, the latter is the
quantum correction. Is this last term which has a clear interpretation in terms of
counting the states inside the system. 

As a cross-check
of this result, \eqref{qentropy}, we take the well-known limit of the Schwarschild case, $H=0$, $r=2m$ and $\beta_{\rm h}=4m$. Then  \eqref{qentropy} simplifies to
\begin{equation}\label{entropySchw}
\begin{split}
    \delta S_{\rm Sch BH}^{\rm q}
     =\frac{A_{r_-}}{4}\left(    \frac{1}{360\pi m^2}\right)\ln{\left(\frac{\Sigma}{\epsilon}\right)}\,,
\end{split}
\end{equation}
recovering\footnote{The second terms in \eqref{entropySchw} differs from the equation (86) in \cite{Solodukhin:1994yz} by a factor of 2. However, it is clear by looking at the process from equation (84) to equation (85) in \cite{Solodukhin:1994yz} that the missing factor 2 in equation (86) is just a typo.} (86) in \cite{Solodukhin:1994yz}. 
On the other hand, if we consider the deSitter case, $m=0$ and $r=H^{-1}$,  in \eqref{qentropy}, we get
\begin{equation}\label{entropycorrectionsdS}
   \delta S_{\rm dS}^{\rm q}
     =  \frac{A_{\rm h}}{4}  \left(    \frac{29 H^2}{180 \pi }\right)\ln{\left(\frac{\Sigma}{\epsilon}\right)}\,.
\end{equation}

In both cases, \eqref{entropySchw} and \eqref{entropycorrectionsdS}, we recover
the entropy to area corrections to the Rindler space in the limits $m\to \infty$ and $H\to 0$ respectively.

One can ask what is the entropy in a shell of width $\delta r$ that an observer, at constant distance from the black hole, measures. One expects that this entropy is related to the generalized mass \eqref{misner}. Using \eqref{qentropy} one gets
\bea
\label{shellS}
\delta S_{\rm shell}^{\rm q}=
  2r \delta r\left(\frac{2H^2}{15 }\right)\ln{\left(\frac{\Sigma}{\epsilon}\right)}\,.
\eea
Notice that \eqref{shellS} is independent of $m$ but not of $H$.
As expected, $\delta S^{\rm q}_{\rm shell}$ vanishes at the limit $H\to 0$.

\section{Final remarks}

 We have shown that the isometries of \eqref{genNewt} and the symmetries of \eqref{quadratelo} are generically not related. When one approaches the Nariai limit the isometries of the spacetime are enhanced while
the solutions to the equation of motion for a massless scalar are not unless one take very large inner radius, or what is equivalent small masses. Only in that case both, the geometry and the Klein-Gordon equation for a massless scalar share the same $SL(2,\mathbb{R})$ symmetry. This is an indication, as in the rotating Nariai/CFT duality \cite{Anninos:2009yc}, the Kerr/CFT \cite{Hartman:2008pb} and the dS/CFT \cite{Strominger:2001pn}, of the possible existence
of a holographic dual to a chiral half of a two-dimensional CFT \cite{Balasubramanian:2009bg}.

\vspace{2cm}

{\bf Acknowledgements}

\noindent P. T. is partially supported by grant PID2019-105614GB-C22, Ministerio de Economía y Competitividad de Espa\~na.

\bibliographystyle{unsrt}
\bibliography{whitehole} 

\end{document}